\documentclass[article]{revtex4-1}
\usepackage{graphicx} 
\usepackage[bb=px]{mathalfa}
\usepackage{xcolor}

\newcommand{\oz}{{\overline z}}
\newcommand{\di}{d}

\newcommand{\E}{\mathbb{E}}
\newcommand{\C}{\mathbb{C}\mathrm{ov}}

\begin{document}

\title{Quantum and Thermal versions of Equilibrium Propagation}

\title{Equilibrium Propagation: the Quantum and the Thermal Cases}

\author{Serge Massar}
\email{Serge.Massar@ulb.be}
\affiliation{Laboratoire d’Information Quantique CP224, Universit\'e libre de Bruxelles (ULB), Av. F. D. Roosevelt 50, 1050 Bruxelles, Belgium}
\author{Bortolo Matteo Mognetti}
\email{Bortolo.Matteo.Mognetti@ulb.be}
\affiliation{
Interdisciplinary Center for Nonlinear Phenomena and Complex Systems CP231, Universit\'e Libre de Bruxelles (ULB), B-1050 Brussels, Belgium}

\begin{abstract}
Equilibrium propagation is a recently introduced method to use and train artificial neural networks in which the network is at the minimum (more generally extremum) of an energy functional. Equilibrium propagation has shown good performance on a number of benchmark tasks. Here we extend equilibrium propagation in two directions. First we show that there is a natural quantum generalization of equilibrium propagation in which a quantum neural network is taken to be in the ground state (more generally any eigenstate) of the network Hamiltonian, with a similar training mechanism that exploits the fact that the mean energy is extremal on eigenstates. Second we extend the analysis of equilibrium propagation at finite temperature, showing that thermal fluctuations allow one to naturally  train the network without having to clamp the output layer during training. We also study the low temperature limit of equilibrium propagation.
\end{abstract}

\date{\today}

\maketitle

\section{Introduction}

Artificial neural networks have achieved impressive results in very disparate tasks, both in science and in everyday life. The bottleneck in the optimization of artificial neural networks is the learning procedure, i.e., the process through which the internal parameters of the model are optimized to accomplish a desired task. The learning procedure used in the best networks today is gradient descent, in which the internal parameters are incrementally changed in order to improve performance, as measured by a cost function. In feed forward networks this procedure can be implemented efficiently, using error backpropagation. In more complex networks it is implemented by backpropagation through time.

Biological systems that learn do not seem to use error backpropagation as the latter cannot be naturally performed by the internal dynamics of the system. Better understanding of biological learning systems could pass through 
 developing learning algorithms in which the two phases of the model (the neuronal and the learning dynamics) can be implemented using similar procedures (or the same circuitry). Such approaches may also be particularly interesting for implementation in analog physical systems, which may lead to improvements in speed or energy consumption.
 
Quantum versions of neural networks and more generally machine learning have attracted much attention recently, as they could offer improved performance over classical algorithms, see e.g. \cite{panella2011neural,rebentrost2018quantum,beer2020training,abbas2021power}. For some reviews, we refer to 
\cite{schuld2014quest,houssein2022machine,cerezo2022challenges}. This field faces significant challenges, such as solving the bottlenecks presented by mapping data from classical to quantum memory, barren plateaus that hinder training \cite{mcclean2018barren}, and of course the difficulty of physically implementing quantum computers.

Many learning algorithms are based on Hebb paradigm, stating that learning processes reinforce synapsis between neurons featuring correlated dynamics \cite{hebb2005organization}, have been extensively studied in the past, e.g., Ref.~\cite{movellan1991contrastive,ackley1985learning,almeida1990learning,pineda1987generalization,xie2003equivalence,poole2022detailed,poole2017chemical}.
Within these algorithms, Equilibrium Propagation (EP)  \cite{SB17}, the focus of the present article,  overcomes some of the limitations of the preceding models. 
Equilibrium Propagation, as described in \cite{SB17}, uses a damped dynamical system that has reached a stationary state. The output of the network is obtained by looking --in the stationary state-- at the state of the output neurons when the input neurons are forced towards the input values (a process known as clamping). To train the network, the output variables are nudged (i.e. clamped) in the desired direction, and the response of the network provides information on the gradients of the cost function. These are then used to  improve the values of the internal parameters. This procedure can be shown to be equivalent to error backpropagation \cite{ernoult2019updates}.

Equilibrium Propagation has been extended in a number of directions. 
Ref.~\cite{L21} proposed symmetric nudging of the output that removes some biases, significantly improving performance. Ref.~\cite{laborieux2022holomorphic} showed how analytical techniques could  be used to exploit large changes in the output, leading to faster convergence of the training dynamics. Equilibrium Propagation in continuous time was studied in \cite{ernoult2020equilibrium}, and using spiking neurons in \cite{martin2021eqspike}.
For a proposal to implement Equilibrium Propagation in electronic systems, see \cite{kendall2020training}.

In this work, we extend the EP algorithm in two directions.
First we show that it can naturally be extended to the quantum setting by taking a quantum neural network to be in an eigenstate of the Hamiltonian. Because eigenstates are extrema of the mean energy, the approach of \cite{SB17} can readily be extended to the quantum case. This mode of operation of a quantum neural network appears to be fundamentaly different from previous proposals. 

Second, we consider (classical or quantum) EP at finite temperature, which was already briefly studied in Ref.~\cite{SB17}. We show that thermal fluctuations provide training 'for free'. More precisely, the gradient of the parameters of the network can be  calculated
 by sampling specific correlations in the free (unclamped) phase without the need to clamp the output.
Moreover, we also provide analytic expressions of the learning and neuronal dynamics in a low-temperature expansion.

The paper is structured as follows: In Sec.~\ref{Sec:EP}, we review the classical implementation of the EP method Ref.~\cite{SB17,L21}. Sec.~\ref{Sec:QEP_gen} is dedicated to the quantum version of EP. In Sec.~\ref{Sec:T} we consider EP at finite temperature. Finally, Sec.~\ref{Sec:Conc} reviews our results and presents ideas for future work.

\begin{figure}[h]
 \includegraphics[height=9cm,width=18.0cm,angle=0]{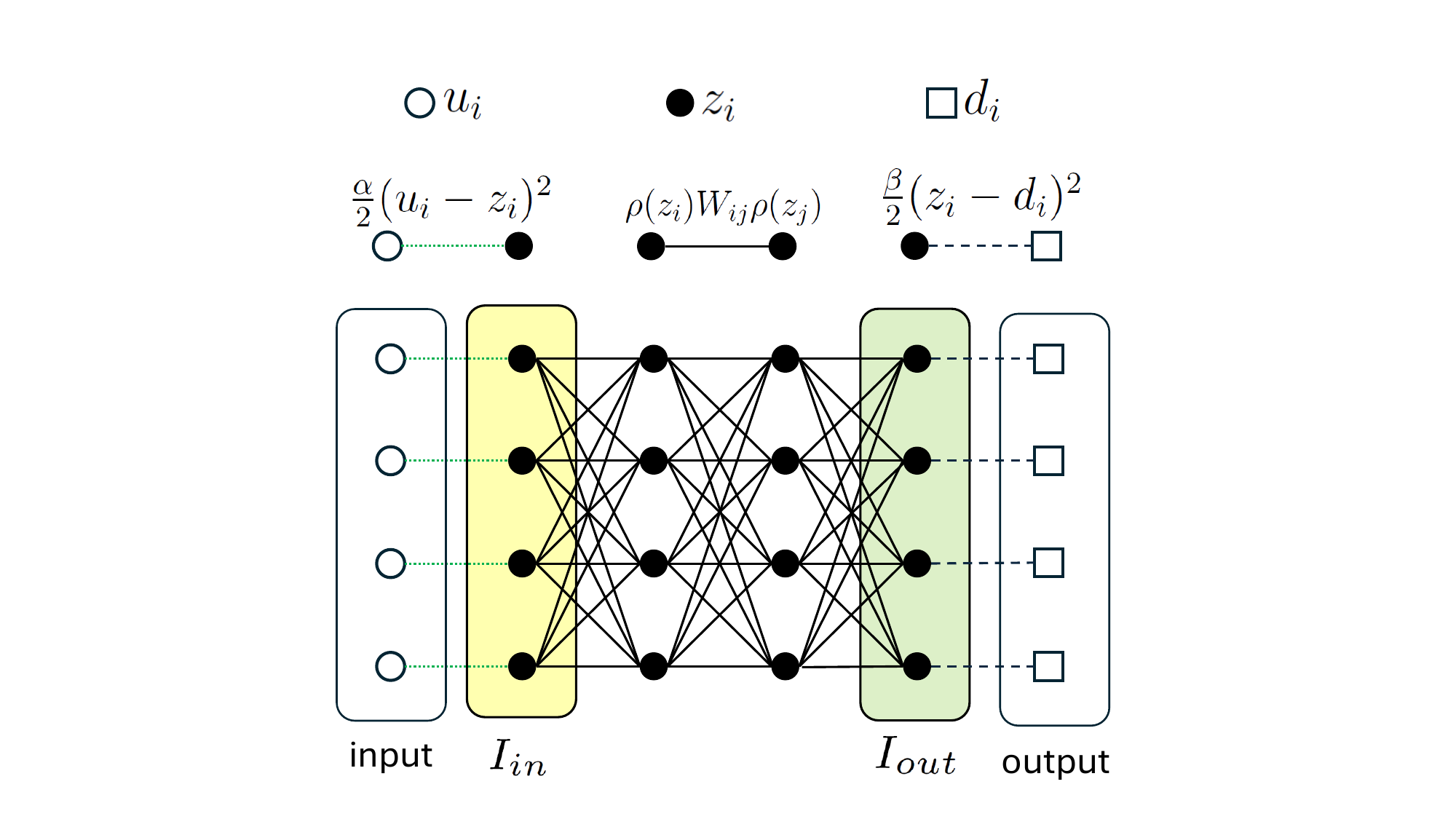}
 \caption{ In Equilibrium Propagation (EP), a neural network, with dynamic variables $z_i$ (represented by black dots) and internal parameters $\{ W_{ij} \}$, is operated at the minimum of an energy functional. During training 
 the input and output layers ($I_{in}$ and $I_{out}$) are clamped to the vectors of the training set ($\{u_i\}$ and $\{ d_i\}$), and the response of the network to the clamping of the output is used to implement gradient descent on the internal parameters $\{ W_{ij} \}$. During normal operations, only the input is clamped, and the response of the output variables is the output of the network.  In the quantum version described in Sec.~\ref{Sec:QEP}, $z_i$ are operators, while $u_i$ and $d_i$ are external forces (for instance magnetic fields). Note that because the network is in equilibrium, information flows through the network in both directions: information on the $u_i$ and $d_i$ is spread out throughout the network. The Equilibrium Propagation method is suitable for general types of topologies (not only multi-layer networks).
}
\label{Fig:fig1}
 \end{figure}

\section{Equilibrium propagation}\label{Sec:EP}

\subsection{Defining the network}

In this section, we review the  Equilibrium Propagation learning algorithm, as introduced in Ref.~\cite{SB17}.
We consider a neural network (Fig.~\ref{Fig:fig1}), comprising an input/output layer and hidden nodes, along with a Hopfield-like potential \cite{hopfield1984neurons}. On each node of the network, we define a neuronal variable, $z_i$, interacting with variables on neighboring nodes through the following potential
\begin{eqnarray}
f_{\theta,\alpha,\beta} (z; u, d) &=& { \alpha \over 2}  \sum_{i \in I_{in}} (u_i-z_i)^2 
+\sum_i h_i(z_i) - {1\over 2} \sum_{i\neq j} W_{ij} \rho(z_i) \rho(z_j) + { \beta \over 2}  \sum_{i\in I_{out}} (z_i-d_i)^2
\label{Eq:f} 
\end{eqnarray}
where  $I_{in}$ and $I_{out}$ are the set of nodes belonging to the input and output layers (see Fig.~\ref{Fig:fig1}), $u$ and $d$ are, respectively, the input and output vectors, $W_{ij}$ quantifies the strength of the connection between node $i$ and node $j$ with $\rho(\cdot )$ an activation function, for instance, $\rho=\tanh$, and $h_i(\cdot )$ the local potential of neuron $i$.
We define by $\theta$ the set of couplings, $\{W_{ij}\}$, along with other parameters defining the functions $h_i$.

To understand how this network  is used, 
let us focus on the situation when $\beta=0$. We aim to adjust the coupling coefficients $W_{ij}$  such that the network carries out a machine learning task, such as classification. For instance, for the MNIST task (classification of handwritten digits) \cite{lecun1998mnist}, the inputs $u_i$ could encode the greyscale values of the image, and there would be $\vert I_{out}\vert = 10$ neurons in the output layers. One of the output neurons should take the value $+1$ (say the eighth neuron, if the input was the digit 8), while the 9 other output neurons should take the value $-1$. Reading out the values of the output neurons will then allow one to determine which handwritten digit was presented at the input. In order to perform this task, we use a set of training examples, for which we know the output, and adjust the network variables $W_{ij}$ so that it performs well on the training set. To evaluate performance, one would then check how the network performs on new images (the test set).

When $\alpha, \beta >0$, the network state, $z$, is clamped to the input and output layer ($u$ and $d$). Accordingly, $u$ and $d$ affect the dynamics of the network (see below). When training the network, we will set $\alpha>0$ and $\beta>0$, and use a dataset constituted by input/output vectors, $\{\{u, d \}\}$. When the network is used on new data (e.g. during testing), $\beta$ is set to $\beta=0$. We will see below that in thermal equilibrium it is not necessary to clamp the output (clamping the input is still necessary), although one still needs to know the desired output $d$.

Note that there is a very large freedom possible in choosing the potential (\ref{Eq:f}).
For instance, for the sake of symmetry, we have used the same form for the input and output coupling, while in \cite{SB17} a linear input coupling was used. These are essentially equivalent. Indeed supposing that 
the input is binary $u_i\in \{-1,+1\}$, then $(z_i - u_i)^2 = -2z_i u_i +(z^2_i +1)$ and the last two terms can (for fixed $\alpha$) be absorbed in $h_i$, yielding a linear coupling. In general, different potentials, or more specifically different clamping strategies, may not perform equally well (see e.g.~\cite{poole2022detailed}).

We define the cost function, $c(z,d)$ as 
\begin{equation}
c(z,d) = {1\over 2} \sum_{i\in I_{out}} \left( z_i  - d_i \right)^2 \, .
\label{Eq:Cost_out}
\end{equation}
The cost function is proportional to the squared Euclidean distance between the output layer and the target vector, and thus measures how much the output differs from the desired output.

\subsection{Using the network}\label{Sec:use}

Let us fix the parameters $\{\theta,\alpha,\beta\}$ and the input/output vectors $\{u,d\}$. We consider that the neuronal dynamics are such that they will tend to minimise the energy. For instance they could be given by  
\begin{eqnarray}
    {d z_i \over d t} &=& -{\partial f_{\theta,\alpha, \beta} (z; u,d) \over \partial z_i } \, .
    \label{Eq:EvolCl}
\end{eqnarray}
The network is used in the steady state, that is when $z= \oz(\alpha,\beta,\theta,u,d)$ with $\oz$ solving the following implicit equation
\begin{eqnarray}
    {\partial f_{\theta,\alpha,\beta}  \over \partial z_i} (\oz; u,d) = 0 \qquad \forall i\ .
\label{Eq:oz}
\end{eqnarray}
Although not essential for the following, we suppose that there is a unique energy minimum.

One of the interests of equilibrium propagation is that one can imagine implementing it in analog systems with very fast dynamics, which reach the steady state solution (Eq.~\ref{Eq:oz}) very fast. Instead, in numerical implementations one needs to integrate the evolution equations (Eq.~\ref{Eq:EvolCl}) which is time consuming.

We denote by $F(\alpha,\beta,\theta, u,d)$ the value of the potential (Eq.~\ref{Eq:f}) evaluated at the steady solution
\begin{eqnarray}
    F(\alpha,\beta,\theta, u,d) = f_{\theta,\alpha,\beta}(\oz; u,d)
    \label{Eq:FT0}
\end{eqnarray}

We denote by $C(\alpha,\beta,\theta, u,d)$ the cost function  (Eq.~\ref{Eq:Cost_out}) evaluated at the steady solution
\begin{equation}
C(\alpha,\beta,\theta, u,d) = c(\oz(\alpha,\beta,\theta, u,d),d)=
{1\over 2} \sum_{i\in I_{out}} \left( \oz(\alpha,\beta,\theta, u,d)  - d_i \right)^2 \, .
\label{Eq:Cost_T0}
\end{equation}

After the training phase, we want the network (with $\beta=0$), when fed by fresh data (e.g., from the test data set), to generate steady state configurations with a small cost. To achieve this we need to train the network. This is carried out 
during t as follows.

During the training phase the parameters of the model $\theta$ are optimized to minimize the cost function through a gradient descent dynamics generated by the cost function
\begin{eqnarray}
\theta \to \theta - \tau \frac{d C}{d\theta} (\alpha,\beta=0 , \theta, u, d ) \, , 
\label{Eq:Learning_gen}
\end{eqnarray}
where $\tau$ is the learning rate. Note that the cost function is evaluated at $\beta=0$ as in the generative/test phase (i.e. the output layer is not clamped to $d$). 

The naive way of implementing the gradient descent (Eq.~\ref{Eq:Learning_gen}) is to successively change each parameter a little bit, let the system reequilibrate, and measure how much the cost has changed. This is slow because it requires as many equilibrations as there are parameters to train. One of the main interests of equilibrium propagation is that one can use only one (better two, see \cite{L21})  equilibrium configurations to learn all the gradients. This is done by clamping the output (i.e. setting $\beta \neq 0$), as we now show.

The key relation which allows this economy of training operations is the equality
\begin{eqnarray}
\frac{d^2 F}{d\theta d\beta} (\alpha, \beta , \theta, u,d) = \frac{d^2 F}{ d\beta d\theta} (\alpha, \beta , \theta, u,d) \ .
\end{eqnarray}

To exploit this, let us compute the derivative of $F(\alpha, \beta , \theta, u,d)$ (Eq.~\ref{Eq:FT0}):
\begin{eqnarray}
\frac{\di F}{\di \beta} (\alpha, \beta , \theta, u,d) &=& \frac{\partial f_{\theta,\alpha,\beta} }{\partial z_i} (\oz; u,d) \frac{\partial \oz_i}{\partial \beta}(\alpha, \beta , \theta, u,d) +  \frac{\partial f_{\theta,\alpha,\beta} }{\partial \beta} (\oz; u,d)  \nonumber\\
&=& 
\frac{\partial f_{\theta,\alpha,\beta} }{\partial \beta} (\oz; u,d)
= C(\alpha , \beta, \theta, u,d) \,
\label{Eq:dE0}
\end{eqnarray}
where we have used Eq.~\ref{Eq:oz}.

Similarly we have
\begin{equation}
\frac{\di F}{\di \theta} (\alpha, \beta , \theta, u,d) =  \frac{\partial f_{\theta,\alpha,\beta} }{\partial \theta} (\oz; u,d)  \ .
\label{Eq:dEdtheta}
\end{equation}
Thus for instance, using the form Eq.~\ref{Eq:f}, we have
\begin{equation}
\frac{\di F}{\di W_{ij}} (\alpha, \beta , \theta, u,d) =  
-\rho(\oz_i) \rho(\oz_j)
 \ .
\label{Eq:dEdtheta-2}
\end{equation}

 The gradient of the cost function (to be used in Eq.~\ref{Eq:Learning_gen}) is calculated by taking the derivative of Eq.~\ref{Eq:dE0}
\begin{equation}
\frac{d C}{d\theta} (\alpha, \beta , \theta, u,d) = \frac{d^2 F}{d\theta d\beta} (\alpha, \beta , \theta, u,d) = \frac{d}{ d\beta}  \frac{\partial f_{\theta,\alpha,\beta} }{\partial \theta} (\oz, u,d)
\label{Eq:Cost_gradient}
\end{equation}
The previous equation states that the derivative of the cost function with respect to a parameter ($\theta$) of the potential $f_{\theta,\alpha,\beta}$ (Eq.~\ref{Eq:f}) is the derivative in the clamping coupling ($\beta$) of the operator conjugated to $\theta$ in $f_{\theta,\alpha,\beta}$. In particular for $W_{ij}$ we find 
\begin{eqnarray}
    \frac{d C}{d W_{ij}} (\alpha, \beta , \theta, u,d) = -\frac{d}{ d\beta} \left[ \rho (\oz_i) \rho(\oz_j)\right] 
    \label{Eq:Cost_gradient_Wij}
\end{eqnarray}
Ref.~\cite{SB17} estimated the right-hand side of Eq.~\ref{Eq:Cost_gradient_Wij} by discretizing the derivative in $\beta$. Ref. \cite{L21} uses a symmetric discretization that reduces biases due to higher-order terms 
\begin{eqnarray}
    \frac{d C}{d W_{ij}} (\alpha, 0 , \theta, u,d) &=& - { \rho (\oz_i^{(+\beta)}) \rho(\oz_j^{(+\beta)}) - \rho (\oz_i^{(-\beta)}) \rho(\oz_j^{(-\beta)})  \over 2 \Delta \beta} 
        \label{Eq:Cost_gradient_Wij_dis}
    \\
    \text{where}\quad \quad
    \oz_i^{(\pm \beta)} &=& \oz_i (\alpha,\pm \beta,\theta, u) \, .
\end{eqnarray}

 In practice, for a pair of input/output vectors $\{u,d\}$ belonging to the training set, the right-hand side of the previous equation is estimated by solving Eq.~\ref{Eq:oz} for two different values of $\beta$. 
 Upon determining how the equilibrium configuration has changed due to the clamping, i.e. determining $\oz_i^{(\pm \beta)}$, one can evaluate the r.h.s. of Eq. \ref{Eq:Cost_gradient_Wij_dis}.
 The coupling parameters $\delta W_{ij}$ are then updated  using Eq.~\ref{Eq:Learning_gen}.

\section{Quantum Equilibrium Propagation}\label{Sec:QEP_gen}

\subsection{Machine learning using quantum networks}

Consider a Hamiltonian consisting of $N$ qubits  of the form
\begin{equation}
\hat H(\beta, \theta, u,  d) = \sum_{i \in I_{in}} u_i  \hat Z_i 
+ \sum_{ij} a_{ij} \hat X_i \hat X_j + b_{ij} \hat Z_i \hat Z_j 
+ {\beta\over 2} \sum_{i \in I_{out}} ( \hat Z_i - d_i  \hat 1 )^2 
\label{Eq:Hthetabeta}
\end{equation}
where $\hat Z$ and $\hat X$ are the Pauli matrices ($\hat Z= \hat \sigma_z$ and $\hat X=\hat \sigma_x$), $I_{in}, \, I_{out}$ ($I_{in}, I_{out} \subset \{1,...,N\}$) are subsets of the qubits constituting the input and output layer, $\hat 1$ the identity operator, and $\theta$ the ensemble of qubit-qubit couplings $\theta=\{ a_{ij},b_{ij}\}$.
Qubits in the input layer are driven by an external magnetic field in the $z$ direction ($u_i$). Qubits in the output layer 
can either be used in an unclamped 
configuration ($\beta=0$) when they should be polarised along the $z$ direction with the desired value ($d_i\in [-1,+1]$), or in a clamped configuration $\beta\neq 0$.

The Hamiltonian Eq.  \ref{Eq:Hthetabeta} is only an example used for illustration.
As already noted in the previous section, it is not necessary to clamp the qubits of the output layer using quadratic terms. For instance, if  $d_i \in \{-1,+1\}$, then the  clamping term becomes linear given that $(\hat Z_i -  d_i \hat 1)^2  = 2 \hat 1 - 2 d_i \hat Z_i $ since the Pauli matrices square to $\hat 1$.
Note also that the Hamiltonian could also involve continuous variables obeying to the canonical commutation rule, $[\hat x,\hat p]=i$. 
Finally note that
we could use different network topologies. For instance, we could arrange the qubits on a 2-dimensional square lattice with nearest-neighbor couplings, with the input and output qubits placed on two opposite edges of the network (in which case most of the $c_{ij}$ and $d_{ij}$ are set to zero).

We define the energy ground state as $\vert \psi_0(u , d, \beta, \theta) \rangle$. Our aim is for the ground state to solve the machine learning task.
 That is we desire that
\begin{equation}
\langle  \psi_0(u , \cdot , \beta=0, \theta)  \vert \hat Z_i \vert \psi_0(u , \cdot , \beta=0, \theta)  \rangle = d_i \quad , \quad i \in I_{out}\ .
\label{Eq:H_classification}
\end{equation}
where we used a dot in the dependency of the ground state to indicate that, for $\beta=0$, $\psi_0$ does not depend on $d$.

Achieving the goal stated by Eq.~\ref{Eq:H_classification} requires minimizing the expectation of the cost function operator (last term of the Hamiltonian, Eq.~\ref{Eq:Hthetabeta}):
\begin{equation}
\hat C (d)= {1\over 2} \sum_{i \in I_{out}} (\hat Z_i - d_i \hat 1)^2 \ .
\label{Eq:Cost_op}
\end{equation}
The expectation of the cost is 
\begin{equation}
\langle C \rangle_0 
= \langle  \psi_0(u , d , \beta, \theta) \vert \hat C(d) \vert \psi_0(u , d , \beta, \theta)  \rangle
\label{Eq:Cost_Q}
\end{equation}
where on the right hand side we have explicitly written out all the dependencies on $(u , d , \beta, \theta) $.
The cost is minimized by adjusting the internal parameters through a gradient descent procedure, as in Eq.~\ref{Eq:Learning_gen}. To this end, we need to evaluate the 
 gradient of the expectation of the cost function in the ground state
 $ \partial  \langle C \rangle_0  / \partial \theta $.
We now show how to efficiently estimate this gradient using 
the quantum variational principle.

\subsection{Quantum Variational Method}\label{Sec:Var}

We recall the variational principle for finding energy eigenstates. 
Let $E$ be the expectation value of the Hamiltonian, viewed as a functional of (unnormalized) states $\vert \psi \rangle$ and their complex conjugate $\langle \psi \vert$:
\begin{equation}
E = \frac{\langle \psi \vert \hat H \vert \psi \rangle}{\langle \psi \vert \psi \rangle}\ .
\label{Epsi}
\end{equation}
The change in $E$ by a variation of $\langle \psi \vert$ is given by
\begin{equation}
\frac{ \delta E}{ \delta \langle \psi \vert } = \frac{ \hat H \vert \psi \rangle}{\langle \psi \vert \psi \rangle}
- \frac{\langle \psi \vert \hat H \vert \psi \rangle}{\langle \psi \vert \psi \rangle^2} \vert \psi \rangle
\ .
\label{Evarpsi}
\end{equation}
A similar expression is found when considering variations in $\vert \psi \rangle$. Eq.~\ref{Evarpsi} can be easily derived using a matrix and vectorial representation, respectively, for $\hat H$ and $\langle \psi \vert $, $\vert \psi \rangle$ and calculating the variation of $E$ with respect to the real and imaginary part of the coefficients defining $\vert \psi \rangle$.
Eq.~\ref{Evarpsi} implies that the extrema of $E$ are energy eigenstates:
\begin{equation}
\frac{ \delta E}{ \delta \langle \psi \vert } = 0  \quad \leftrightarrow \quad
H \vert \psi \rangle = \lambda \vert \psi \rangle  \quad , \quad \lambda = \frac{\langle \psi \vert H \vert \psi \rangle}{\langle \psi \vert \psi \rangle}\ .
\label{Eq:Hextrem}
\end{equation}

\subsection{Quantum Equilibrium Propagation}\label{Sec:QEP}

In the following, we show how to efficiently estimate the gradient of the cost function $\partial \langle C \rangle_0 / \partial \theta $ using only a few ground states. The reasoning is similar to that given in section \ref{Sec:use}.
Let 
\begin{equation}
E_0(\beta,\theta,u,d) = \frac{\langle \psi_0 (u,d,\theta,\beta)\vert \hat H(u,d,\theta,\beta) \vert \psi_0(u,d,\theta,\beta) \rangle}{\langle \psi_0(u,d,\theta,\beta) \vert \psi_0(u,d,\theta,\beta)\rangle}
\end{equation}
be the energy of the ground state as a function of $u$, $d$, $\theta$, and $\beta$. 

Similar to the classical case, the cost function is given by the derivative of $E_0$ with respect to $\beta$. Indeed, using Eq.~\ref{Eq:Hthetabeta} and without making explicit the dependency on $\{u,d,\theta,\beta\}$ to keep the notation compact, we find 
\begin{eqnarray}
\frac{\partial E_0  }{ \partial \beta}(\beta,\theta,u,d)
&=&
\frac{ \delta E_0}{ \delta \langle \psi_0  \vert } \frac{\partial \langle \psi_0 \vert   }{ \partial \beta}
+
\frac{ \delta E_0}{ \delta \vert \psi_0 \rangle} \frac{\partial \vert\psi_0 \rangle   }{ \partial \beta}
+\frac{\langle \psi_0 \vert \frac{ \partial \hat H}{\partial \beta} \vert \psi_0\rangle}{\langle \psi_0 \vert \psi_0\rangle}\nonumber\\
&=& \frac{\langle \psi_0 \vert \frac{ \partial \hat H}{\partial \beta} \vert \psi_0\rangle}{\langle \psi_0 \vert \psi_0\rangle} = \langle C \rangle_0  \, \, .
\label{Eq:dEdbe_Q}
\end{eqnarray}
In the second line, we have used the fact that $\vert \psi_0\rangle$ is an eigenstate, hence it extremises $E_0$ (see Eq.~\ref{Eq:Hextrem}), and therefore the first two terms on the right-hand side of the first equation vanish.
 We have also used the fact that the derivative of the Hamiltonian with respect to $\beta$ is the cost operator $\hat C$ (Eq.~\ref{Eq:Cost_op}). Thus  Eq.~\ref{Eq:dEdbe_Q} generalizes Eq.~\ref{Eq:dE0} to quantum systems.
 
 Similarly we also find  
\begin{eqnarray}
\frac{\partial E_0  }{ \partial \theta}(\beta,\theta,u,d)
&=& \frac{\langle \psi_0 \vert \frac{ \partial \hat H}{\partial \theta} \vert \psi_0\rangle}{\langle \psi_0 \vert \psi_0\rangle} \, \, .
\label{Eq:dEdbe_Q2}
\end{eqnarray}
(Note that Eqs. \ref{Eq:dEdbe_Q} and \ref{Eq:dEdbe_Q2} are in fact immediate applications of first order perturbation theory).

We then follow what has already been done in Sec.~\ref{Sec:EP}. In particular, using the identity
\begin{equation}
\frac{\partial^2 E_0 }{\partial \theta \partial \beta}
(\beta,\theta,u,d)
=
\frac{\partial^2 E_0  }{\partial \beta\partial \theta }
(\beta,\theta,u,d)
\label{Eq:Identity}
\end{equation}
we find 
\begin{equation}
\frac{\partial C }{\partial \theta }
(\beta,\theta,u,d)
=
\frac{\partial  }{\partial \beta } \frac{\langle \psi_0 \vert \frac{ \partial \hat H}{\partial \theta} \vert \psi_0\rangle}{\langle \psi_0 \vert \psi_0\rangle}
\label{Eq:Identity2}
\end{equation}

Therefore we can estimate the derivative of the cost function with respect to $\theta$ by computing the change in the expectation value of 
$\partial {\hat{H}(\theta)}/{\partial \theta}$ in the ground state when we let $\beta$ vary: 
\begin{equation}
\frac{\partial \langle C \rangle_0^{(\beta=0)} }{\partial \theta }
\approx 
\frac{
 \frac{
 \langle \psi_0\vert \frac{ \partial \hat H}{\partial \theta} \vert \psi_0 \rangle}{\langle \psi_0 \vert \psi_0 \rangle}^{(+\beta)}
-  \frac{\langle \psi_0\vert \frac{ \partial \hat H}{\partial \theta} \vert \psi_0 \rangle}{\langle \psi_0 \vert \psi_0 \rangle}^{(-\beta)}
}
{2 \Delta \beta}\ .
\label{Eq:QEP}
\end{equation}
where the superscripts $(\beta=0), (+\beta),(-\beta)$ indicate for the value to which $\beta$ is set in the Hamiltonian.

Thus for instance, going back to the explicit example Hamiltonian Eq. (\ref{Eq:Hthetabeta}), we have
\begin{equation}
\frac{\partial \langle C \rangle_0^{(\beta=0)} }{\partial a_{ij} }
\approx 
\frac{
 \frac{
 \langle \psi_0\vert \hat X_i \hat X_j  \vert \psi_0 \rangle}{\langle \psi_0 \vert \psi_0 \rangle}^{(+\beta)}
-  \frac{\langle \psi_0\vert \hat X_i \hat X_j  \vert \psi_0 \rangle}{\langle \psi_0 \vert \psi_0 \rangle}^{(-\beta)}
}
{2 \Delta \beta}\ .
\label{Eq:QEP-example}
\end{equation}

In practice we would proceed as follows. Construct the states 
$\vert \psi_0 \rangle^{(\pm\beta)}$. 
Measure all the operators $\hat X_i$ in this state. (They can be measured simultaneously since they commute). Repeat this enough times to get reliable estimates for the expectation values in Eq.~\ref{Eq:QEP-example}. Use this estimate  to update the parameters $a_{ij}$. Note that to update the parameters $b_{ij}$ one would need to repeat the procedure, since the operators $\hat X_i$ and $\hat Z_i$ do not commute, and hence cannot be measured simultaneously.

\section{Equilibrium propagation at finite temperature}\label{Sec:T}

\subsection{General setting}

In Ref.~\cite{SB17}, Equilibrium Propagation was generalised to the stochastic framework, corresponding to the finite temperature situation. Here we revisit the finite temperature setting.

We suppose that the configurations of the system are distributed according to  the Boltzmann distribution \cite{van1992stochastic}
\begin{eqnarray}
    p_{\alpha,\beta,\theta,u,d,T} (z) &=& e^{ {F(\alpha,\beta,\theta,u,d,T) \over T} - {f_{\theta,\alpha,\beta} (z; u,d) \over T} }
    \label{Eq:Boltzmann}
\end{eqnarray}
where we have defined the 'free energy' $F(\alpha,\beta,\theta,u,d,T)$ as \cite{SB17}
\begin{eqnarray}
  e^{- {F(\alpha,\beta,\theta,u,d,T) \over T} } = \int \di z \, e^{- {f_{\theta,\alpha,\beta} (z; u, d) \over T } } \, .
\label{Eq:FthetaT}  
\end{eqnarray}
The free energy $F(\alpha,\beta,\theta,u,d,T)$ generalizes equation \ref{Eq:FT0} to the finite temperature case, and to indicate this we add the explicit temperature dependence to $F$.

In the quantum case, we would define the free energy as
\begin{eqnarray}
  e^{- {F(\alpha,\beta,\theta,u,d,T) \over T} } &=& \text{Tr} \  e^{- {\hat {H}(\beta, \theta, u, d) \over T } } \, .
\label{Eq:FthetaT-quantum}  
\end{eqnarray}
In what follows we consider for definiteness the classical case and the energy functional Eq.~\ref{Eq:f}, but all expressions generalize to the quantum case and to other energy functionals.

In a physical system, the dynamics of $z_i$ would be given by a stochastic equation, such as the Langevin equation \cite{van1992stochastic}
\begin{eqnarray}
    {\di z_i \over \di t} &=& -{\di f_{\theta,\alpha,\beta} (z,u,d) \over \di z_i } +\sqrt{2 T} \eta_i(t) \, ,
\label{Eq:Langevin}
\end{eqnarray}
where  $\eta_i(t)$ are normally distributed random variables, $\langle \eta_i(t)\eta_j(t') \rangle = \delta_{i,j} \delta(t-t')$, and $T$ the temperature of the system, such that the solutions of the stochastic process follow the Boltzmann distribution. In the quantum case one could replace Eq.~\ref{Eq:Langevin} by a Lindblad equation.

The expectation of the cost function Eq. \ref{Eq:Cost_out} reads as follows \cite{SB17}
\begin{eqnarray}
   \langle C \rangle_T
    &=& \mathbb{E}_{\alpha,\beta,\theta,u,d,T} \left[
    c (z,d) \right]
    \label{Eq:Cost_T}
\nonumber
\end{eqnarray}
 where for simplicity in the notation $ \langle C \rangle_T$ we indicate that we evaluate at finite temperature, and have not written explicitly the dependence on $\alpha,\beta,\theta,u,d$.
 The average in Eq.~\ref{Eq:Cost_T} is calculated using the Boltzmann distribution (Eq.~\ref{Eq:Boltzmann}) as follows
\begin{eqnarray}
\mathbb{E}_{\alpha,\beta,\theta,u,d,T} \left[ A(z) \right] = \int \di z\,  p_{\alpha,\beta,\theta,u,d,T} (z)  A(z)  
\label{Eq:averageT}
\end{eqnarray}
where $A(z)$ is a generic function of the network's state.

We now compute the derivatives of the free energy.
From Eq.~\ref{Eq:FthetaT} and Eq.~\ref{Eq:f} we find that 
\begin{eqnarray}
 {d F \over d W_{ij} } (\alpha,\beta,\theta,u,d,T) 
 &=& - \mathbb{E}_{\alpha,\beta,\theta,u,d,T} \left[  \rho(z_i) \rho(z_j) \right]
 \label{Eq:dFdW}
 \\
 {\di F \over \di \beta } (\alpha,\beta,\theta,u,d,T) 
 &=&  \mathbb{E}_{\alpha,\beta,\theta,u,d,T} \left[ {1\over 2} \sum_{i\in I_{out}} (y_i -d_i)^2 \right] 
 \nonumber \\
 &=& \langle C \rangle_T \, ,
\label{Eq:dFdbeta}
\end{eqnarray}

Using Eqs.~\ref{Eq:dFdbeta}, \ref{Eq:FthetaT}, \ref{Eq:averageT}, and \ref{Eq:Boltzmann}, we can calculate the gradient of the cost function to be used in the learning dynamics (Eq.~\ref{Eq:Learning_gen}),
\begin{eqnarray}
 {d \langle C \rangle_T \over d W_{ij}} (\alpha,\beta=0,\theta,u,d,T) &=&   { d^2 F_T \over d \beta d W_{ij} } (\alpha,\beta=0,\theta,u,d,T) 
 \label{Eq:Cost_gradient_T} \\
 &=& - {d \over d \beta } \mathbb{E}_{\alpha,\beta=0,\theta,u,d,T} \left[  \rho(z_i) \rho(z_j) \right]
 \label{Eq:LearningT_clamp}
\\
&=&  \frac{1}{T}\C_{\alpha,\beta=0,\theta,u,\cdot,T} \left[\rho(z_i) \rho(z_j),   c(z,d)  \right] 
\label{Eq:LearningT_corr}
\end{eqnarray}
where $\C [X,Y ] = \E [ X Y ] - \E [ X ]\E [ Y ]$. 

Eq.~\ref{Eq:LearningT_corr} can be obtained by taking the derivative of the free energy definition Eq.~\ref{Eq:FthetaT}.
An alternate proof of the equivalence between Eq.~\ref{Eq:LearningT_clamp} and Eq.~\ref{Eq:LearningT_corr} can be obtained using statistical reweighting. Specifically, using the definition of the Boltzmann distribution (Eq.~\ref{Eq:Boltzmann}), we can write 
\begin{eqnarray}
\E_{\alpha,\beta,\theta,u,d,T}\left[ \rho(z_i) \rho(z_j)  \right] &=& { {\E_{\alpha,\beta=0,\theta,u,d,T}\left[ \rho(z_i) \rho(z_j) e^{-{\beta\over  T} c(z,d) } \right]} \over \E_{\alpha,\beta=0,\theta,u,d,T}\left[  e^{-{\beta\over T} c(z,d) }  \right] }
\label{Eq:reweigh} \\
& \approx & 
- \frac{\beta}{T} \C_{\alpha,\beta=0,\theta,u,\cdot,T} \left[ \rho(z_i) \rho(z_j) , { c(z,d) \ }   \right]
\label{Eq:LearningT_clamp_T0}
\end{eqnarray}
where the second equality has been taken in the small $\beta$ limit. Inserting Eq.~\ref{Eq:reweigh} into Eq.~\ref{Eq:LearningT_clamp} yields Eq.~\ref{Eq:LearningT_corr}.

Similarly to what was done in Eq.~\ref{Eq:Cost_gradient_Wij_dis},
Eq.~\ref{Eq:LearningT_clamp} suggests a learning dynamics in which the derivative in $\beta$ in the r.h.s.~of Eq.~\ref{Eq:LearningT_clamp} is evaluated by clamping the system to $d$ using two different values of $\beta$ while sampling $\rho(z_i) \rho(z_j)$ \cite{SB17,L21}
\begin{eqnarray}
\delta W_{ij} 
&=& -\tau { \E_{\alpha,+\beta,\theta,u,d,T}\left[ \rho(z_i) \rho(z_j)  \right] - \E_{\alpha,-\beta,\theta,u,d,T}\left[ \rho(z_i) \rho(z_j)  \right] \over 2 \beta } \, .
\label{Eq:LearningT_clamp_T0_2}
\end{eqnarray}

Instead, Eq.~\ref{Eq:LearningT_corr} suggests a learning dynamics in which the system is not clamped to $\{d\}$ (implying that the Boltzmann distribution is not a function of $d$), rather, $\{\delta W_{ij}\}$ are calculated by sampling the covariance between $\rho(z_i) \rho(z_j)$ and the cost function $C(z,d)$
with $\beta=0$
\begin{eqnarray}
\delta W_{ij} 
&=& -\tau \frac{1}{T} \C_{\alpha,\beta=0,\theta,u,d,T} \left[\rho(z_i) \rho(z_j),   c(z,d)  \right] \, .
\label{Eq:LearningT_corr_T0}
\end{eqnarray}
That is, the thermal fluctuations directly provide us with the desired correlations between cost and variables conjugate to $W_{ij}$, without having to clamp the output variables.

\subsection{Low-temperature expansion}\label{Sec:EPsmallT}

For $T=0$, the Boltzmann distribution is a delta function peaked at the minimum of $f_{\theta,\alpha,\beta}(z)$ (Eq.~\ref{Eq:f}), $p_{\alpha,\beta,\theta,u,d,T=0}(z) \sim \delta\left(z-\oz(\alpha,\beta,\theta,u,d) \right)$ with $\oz$ defined in Eq.~\ref{Eq:oz}. Consequently the right-hand side of Eq.~\ref{Eq:LearningT_corr} becomes indeterminate. We now analyse the $T\to 0$ limit and   show that it  is well-defined.

We  consider a low-temperature expansion of the theory developed in the previous section. In doing so, we provide deterministic learning equations that could be  employed in networks for which the Hessian (defined below) and its inverse  is known \cite{stern2024physical}. Note however that the Hessian will depend on the input $u$ and  output $d$, and would need to be reevaluated for each example from the training set.

We take the $T\to 0$ limit of $F(\alpha, \beta, \theta,u,d,T)$ (Eq.~\ref{Eq:FthetaT}) by considering the development of $f_{\theta,\alpha,\beta}(z;u,d)$ (in the right-hand side of Eq.~\ref{Eq:FthetaT}) around its stationary point $\oz$ (Eq.~\ref{Eq:oz}):
\begin{eqnarray}
F_T(\alpha, \beta, \theta,u,d,T) &=& - T \log \int \di z \  e^{- {f_{\theta,\alpha,\beta} (\oz; u,d)\over T} -{1 \over 2 T} H_{ij} \Delta z_i \Delta z_j+O\left( \Delta z^3 \over T\right)}
\label{Eq:F_sp1}
\end{eqnarray}
where $\Delta z_i=z_i - \oz_i$ and $H$ is the Hessian matrix
\begin{eqnarray}
H_{ij} (\alpha, \beta, \theta, u,d) &=& { \partial^2 f_{\theta,\alpha,\beta} \over \partial z_i \partial z_j } (\oz (\alpha, \beta, \theta, u,d); u, d) \, .
\label{Eq:Hessian}
\end{eqnarray}
The explicit expression of the Hessian matrix for the energy $f_{\theta,\alpha,\beta}$ given in Eq.~\ref{Eq:f} is given in Appendix~\ref{App:Hessian}.

By defining $\kappa_i = \Delta z_i / \sqrt{T}$ in Eq.~\ref{Eq:F_sp1}, we calculate the first correction in $T$ to $F(\alpha,\beta,\theta,u,d,T=0)$ given in Eq.~\ref{Eq:FT0}:
\begin{eqnarray}
F(\alpha,\beta,\theta,u,d,T) &=& -T \log \int \di \kappa \cdot T^{D/2} e^{-  {f_{\theta,\alpha,\beta} (\oz; u, d)\over T} -{1\over 2} \sum_{i,j} H_{ij} \kappa_i \kappa_j+O( \kappa^3 \sqrt{T})}
\nonumber
\\
&=& f_{\theta,\alpha,\beta} (\oz; u, d) - {T D\over 2} \log T  - T \log \left[ (2 \pi )^{D/2} (\det   H )^{-1/2}\right]
+O\left( T^2\right)
\nonumber
\\
&=& 
f_{\theta,\alpha,\beta} (\oz; u, d) - {T D\over 2} \log \left( 2 \pi T\right)  + { T \over 2 } \mathrm{Tr} \log H
+O\left( T^2\right)  
\label{Eq:FtheSP}
\end{eqnarray}
where $D$ is the dimension of $H$. 

Using the previous expression we calculate $d F/d W_{ij}$ as follows
\begin{eqnarray}
{\di F \over \di W_{ij}} (\alpha,\beta,\theta,u,d,T) &=& {\partial f_{\theta,\alpha,\beta}  \over \partial W_{ij}}  + \sum_a {\partial f_{\theta,\alpha,\beta}  \over \partial \oz_a} {\partial \oz_p \over \partial W_{ij}} + {T \over 2 } \mathrm{Tr}\left[ H ^{-1} {d  H \over \di W_{ij}} \right]+O\left( T^2\right) 
\nonumber\\
&=&-\rho(\oz_i) \rho(\oz_j) +{T\over 2} \sum_{a,b}
\left[ H^{-1}_{ab} {\partial H_{ab} \over \partial W_{ij}} + \sum_c H^{-1}_{ab} {\partial H_{ab} \over \partial \oz_c} {\partial \oz_c \over \partial W_{ij}} \right] +O\left( T^2\right) 
\nonumber\\
 &=&-\rho(\oz_i) \rho(\oz_j) +{T\over 2} \sum_{a,b}
H^{-1}_{ab}
\left[ {\partial H_{ab} \over \partial W_{ij}} + \sum_c  V^{3}_{abc} {\partial \oz_c \over \partial W_{ij}} \right] +O\left( T^2\right) 
\, ,
\label{Eq:dFdW-2}
\end{eqnarray}
where we have defined 
\begin{eqnarray}
    V^n_{a_1 \cdots a_n} &=& {\partial^n f_{\theta,\alpha,\beta} \over \partial \oz_{a_1} \cdots \partial \oz_{a_n}}(\oz;u,d) \,.
\end{eqnarray}
In Appendix~\ref{App:dFdW} we explicitize Eq. \ref{Eq:dFdW-2} 
 using the Hessian matrix given in App.~\ref{App:Hessian}. Moreover, App.~\ref{App:dFdW} also shows how Eq.~\ref{Eq:dFdW-2} can be derived by developing the right-hand side of Eq.~\ref{Eq:dFdW}. In the same appendix, we also derive the following expression of the derivative of the saddle point solution $\oz$ with respect to a parameter of the network $\theta$  
\begin{eqnarray}
    {\partial \oz_i \over \partial \theta}= -\sum_j H^{-1}_{ij} {\partial^2 f_{\theta,\alpha,\beta}\over \partial \oz_j \partial \theta}(\oz;u,d)  \, .
    \label{Eq:dozdtheta}
\end{eqnarray}
Finally, using Eq.~\ref{Eq:dozdtheta} in Eq.~\ref{Eq:dFdW-2} allows calculating the gradient of $F$ in terms of the Hessian matrix $H$  and the saddle point solution $\oz$.

The gradient of the cost function, to be used in the learning equation (Eq.~\ref{Eq:Learning_gen}), is calculated from the second derivative of $F$, $d^2 F/ d W_{ij} d \beta$ (Eq.~\ref{Eq:Cost_gradient_T}).
Eq.~\ref{Eq:dFdW-2} can be used in the learning equation (Eq.~\ref{Eq:Learning_gen}) as done in Eq.~\ref{Eq:LearningT_clamp} or Eq.~\ref{Eq:LearningT_clamp_T0} by discretizing the derivative in $\beta$ and calculating the stationary point for two clamped networks ($\oz^{(\beta)}$ and $\oz^{(-\beta)}$). Alternatively, one can take the derivative of Eq.~\ref{Eq:dFdW-2} with respect to $\beta$. At the leading order, we find (if $\rho'(z)=\partial \rho(z)/\partial z$)
\begin{eqnarray}
{\di^2 F \over \di W_{ij} \di \beta} (\alpha,\beta=0,\theta,u,d,T)
 &=& \rho'(\oz^{(0)}_i) \rho(\oz^{(0)}_j) {\partial \oz_i \over \partial \beta}\Bigg{\vert}_{\beta=0} + \rho(\oz^{(0)}_i) \rho'(\oz^{(0)}_j) {\partial \oz_j \over \partial \beta}\Bigg{\vert}_{\beta=0} + O(T)
\nonumber \\
&=& \rho'(\oz^{(0)}_i) \rho(\oz^{(0)}_j) \sum_{a\in I_{out}} (\oz^{(0)}_a - d_a) (H_{\beta=0})^{-1}_{ia} 
 \nonumber \\
&& 
\quad + \rho(\oz^{(0)}_i) \rho'(\oz^{(0)}_j) \sum_{a\in I_{out}} (\oz^{(0)}_a -d_a) (H_{\beta=0})^{-1}_{j a} + O\left( T \right)
\label{Eq:dFdWdbe}
\end{eqnarray}
where in the second equality we have used Eq.~\ref{Eq:dozdtheta}. In App.~\ref{App:d2FdWdb} we explicitly report the $O(T)$ contribution.
Notice that Eq.~\ref{Eq:dFdWdbe} is well defined in the $T=0$ limit. In appendix \ref{App:d2FdWdb} we rederive this result by expanding the right-hand side of Eq.~\ref{Eq:LearningT_corr} (which is indeterminate for $T=0$).

So far, we have discussed corrections in $T$ of the learning dynamics. The neuronal dynamics, and in particular the values of $z$ defined on the output layer, are also altered by corrections in $T$. In particular, using Eq.~\ref{Eq:DevSP}, we find 
\begin{eqnarray}
    \E_{\alpha,\beta,\theta,u,d,T}[z_a] &=& \oz_a-{T\over 2} \sum_{b,c,d} V^3_{bcd} H^{-1}_{ab} H^{-1}_{cd} + O(T^2) \, .
\end{eqnarray}

\section{Conclusions}\label{Sec:Conc}

Hebbian-like learning rules \cite{movellan1991contrastive,ackley1985learning,almeida1990learning,pineda1987generalization,xie2003equivalence,poole2022detailed,poole2017chemical,L21,SB17} are being considered as an alternative to error backpropagation algorithms. Their interest relies on the fact that the training phase (backward pass, in which the parameters of the network $\theta$ are optimized) and the neuronal dynamics (forward pass, in which the network state $z$ is updated) can be implemented using a single circuitry or set of operations. This aspect makes them suitable for neuromorphic implementations. Moreover, Hebbian-like learning rules are being used to assess plausible learning dynamics in biological systems.  
 The Equilibrium Propagation (EP) learning algorithm introduced in Ref.~\cite{L21} can overcome some of the problems encountered in other Hebbian methodologies \cite{movellan1991contrastive,ackley1985learning,almeida1990learning,pineda1987generalization,xie2003equivalence,poole2022detailed,poole2017chemical}. Moreover, recent contributions, such as symmetric clamping of the output \cite{SB17} or holomorphic EP \cite{laborieux2022holomorphic},
further strengthen the method. 

Our  contributions concern the extension of the EP algorithm to quantum systems, as well as the use of EP at finite temperature. However we have only presented the principles of the methods, but not yet studied their performance numerically on examples.

In the quantum case, a first important question to address in future work is how to reach an energy eigenstate. Indeed reaching the ground state of a complex Hamiltonian is a notoriously difficult problem. However here we need to be in an eigenstate, but not necessarily the ground state. For this reason the adiabatic quantum algorithm \cite{farhi2000quantum,albash2018adiabatic} (in which the Hamiltonian is gradually changed from an easy Hamiltonian to the desired one) may be useful. Indeed for the present application, avoided crossings at which the adiabatic condition does not hold are not necessarily a problem, as the state will continue being an eigenstate, albeit not the same one. Another method which could be used is algorithmic cooling, in which the quantum system is gradually brought to a low temperature state. As section \ref{Sec:T} shows, one does not need to be in the ground state, only at a sufficiently low temperature state. Another issue that needs to be confronted in the quantum case is the possibility of barren plateaus which would make the gradients  very small. Operating the quantum network near a phase transition, where the system is very sensitive to external perturbations (such as clamping of the output) may help resolve this problem.

In the thermal case, the fact that gradients can be obtained directly from thermal fluctuations is encouraging, and could be very useful in physical implementations of Equilibrium Propagation. However these correlations could be small and difficult to measure. Moreover the performance of Equilibrium Propagation is expected decrease with increasing temperature. The impact of finite temperature on Equilibrium Propagation needs to be studied in detail.

We hope that our work will encourage further investigations of Equilibrium Propagation. 

\begin{acknowledgements} S.M.~would like to thank Guillaume Pourcel for introducing him to Equilibrium Propagation and stimulating his interest in the topic.
Both authors would like to thank Dimitri Vanden Abeele for useful discussions. 
\end{acknowledgements}

\appendix 

\section{Calculations supporting Sec.~\ref{Sec:EPsmallT}}

\subsection{Calculation of the Hessian matrix}\label{App:Hessian}

By using Eq.~\ref{Eq:f} in Eq.~\ref{Eq:Hessian} we find
(if $\rho'(z)=\partial \rho(z)/\partial z$ and $\rho''(z)=\partial^2 \rho(z)/\partial z^2$)
\begin{eqnarray}
H_{ij} &=& t_i \delta_{ij} -\rho'(\oz_i) \rho'(\oz_j) W_{ij} 
\label{Eq:H} \\
t_i &=& \left\{ 
\begin{array}{ll}
\alpha 
-\sum_p W_{ip} \rho''(\oz_i) \rho(\oz_p) & z_i \in \{I_{in}\}
\\
h''(\oz_i) -\sum_p W_{ip} \rho''(\oz_i) \rho(\oz_p)&  z_i \notin \{I_{in},I_{out}\}
\\
\beta 
-\sum_p W_{ip} \rho''(\oz_i) \rho(\oz_p) & z_i \in \{ I_{out}\}
\end{array}
\right.
\nonumber
\end{eqnarray}

\subsection{Calculation of $d F/d W_{ij}$}\label{App:dFdW}

In this section, we derive an expression of $d F/d W_{ij}$ by developing the right-hand side of Eq.~\ref{Eq:dFdW}.
We then further develop Eq.~\ref{Eq:dFdW} using the expression of the Hessian matrix (Eq.~\ref{Eq:H}) and show the consistency of the two results.

The development in $\Delta k$ that led to Eq.~\ref{Eq:FtheSP} can be easily adapted to the calculation of the expectation value of an arbitrary observable $\chi (z)$. In particular, we find 
\begin{eqnarray}
\E_{\alpha,\beta,\theta,u,d,T} \left[ \chi \right] = \chi(\oz) 
+ {T\over 2} \left[ \sum_{a,b} { \partial^2 \chi(\oz) \over  \partial z_a \partial  z_b } H^{-1}_{ab} 
-\sum_{a,b,c,d} V^{3}_{abc} { \partial \chi(\oz) \over \partial z_d}  H^{-1}_{ab} H^{-1}_{cd}  \right] + O\left( T^2\right)\ .
\label{Eq:DevSP}
\end{eqnarray}
The previous equation can be used to estimate the right-hand side of Eq.~\ref{Eq:dFdW} as follows: 
\begin{equation}
    - \mathbb{E}_{\alpha,\beta,\theta,u,d,T} \left[  \rho(z_i) \rho(z_j) \right] = -\rho (\oz_i) \rho (\oz_j) + T \Lambda_1 + O(T^2)
    \nonumber
\end{equation}
with
\begin{eqnarray}
\Lambda_1
&=&
 - {1\over 2 }\Bigg[ \rho''(\oz_i) \rho(\oz_j) H^{-1}_{ii} + \rho(\oz_i) \rho''(\oz_j) H^{-1}_{jj}
 +2 \rho'(\oz_i) \rho'(\oz_j) H^{-1}_{ij}\Bigg] 
\nonumber \\
&& + {1\over 2} \sum_{a,b,c} V^3_{abc} H^{-1}_{bc} \Bigg[ \rho'(\oz_i) \rho(\oz_j) H^{-1}_{ia}  +  \rho(\oz_i) \rho'(\oz_j) H^{-1}_{ja} 
\Bigg]\ .
\label{Eq:dFdW-SP1}
\end{eqnarray}
We now show that Eq.~\ref{Eq:dFdW-SP1} is consistent with Eq.~\ref{Eq:dFdW-2}.
Using Eq.~\ref{Eq:H}, we calculate the second term in the right hand side of Eq.~\ref{Eq:dFdW-2} as follows 
\begin{eqnarray}
\sum_{a,b} H^{-1}_{ab} {\partial H_{ab} \over \partial W_{ij}} &=& 2 H^{-1}_{ij} {\partial H_{ij} \over \partial W_{ij}}+H^{-1}_{ii} {\partial H_{ii} \over \partial W_{ij}}+H^{-1}_{jj} {\partial H_{jj} \over \partial W_{ij}}
\nonumber \\
&=& -2\rho'(\oz_i) \rho'(\oz_j) H^{-1}_{ij}-\rho''(\oz_i) \rho(\oz_j) H^{-1}_{ii}
-\rho(\oz_i) \rho''(\oz_j) H^{-1}_{jj} \ .\quad\ 
\label{Eq:SuppA}
\end{eqnarray}
To unwrap the last term of the r.h.s.~of Eq.~\ref{Eq:dFdW-2}, we need to calculate $\partial \oz/\partial W_{ij}$. This can be done by deriving the saddle-point condition (Eq.~\ref{Eq:oz}) w.r.t.~$W_{ij}$. In particular, we find 
\begin{eqnarray}
{\di \over \di W_{ij}} {\partial f_{\theta,\alpha,\beta}(\oz; u,d)  \over \partial \oz_k} = {\partial^2 f_{\theta,\alpha,\beta}(\oz; u,d) \over \partial \oz_k \partial W_{ij} } + {\partial^2 f_{\theta,\alpha,\beta}(\oz; u,d) \over \partial \oz_k \partial \oz_a } {\partial \oz_a \over \partial W_{ij}}  = 0  \qquad \forall k
\label{Eq:doz_app}
\end{eqnarray}
from which we derive 
\begin{eqnarray}
{\partial \oz_k \over \partial W_{ij}} &=& \sum_a H^{-1}_{ka}{\partial [\rho(\oz_i) \rho(\oz_j)] \over \partial \oz_a}
\nonumber \\
&=& H^{-1}_{ki} \rho'(\oz_i) \rho(\oz_j) +H^{-1}_{kj} \rho(\oz_i) \rho'(\oz_j) \ .
\label{Eq:SuppB}
\end{eqnarray}
Using Eqs.~\ref{Eq:SuppA} and \ref{Eq:SuppB} in Eq.~\ref{Eq:dFdW-2} proves that the right hand sides of Eq.~\ref{Eq:dFdW-SP1} and \ref{Eq:dFdW-2} are equal.

\subsection{Calculation of $d^2 F /d W_{ij} d \beta$}\label{App:d2FdWdb}

In the first part of this section, we prove that Eq.~\ref{Eq:dFdWdbe} can be obtained by expanding the right-hand side of Eq.~\ref{Eq:LearningT_corr_T0}. To do so, we use \ref{Eq:DevSP}. Using such an expression, we have that 
\begin{eqnarray}
\frac{1}{T} \C_{\alpha,\beta=0,\theta,u,\cdot,T} \left[\rho(z_i) \rho(z_j),  { C(z,d) } \right] =   \sum_{a,b} {\partial \rho(\oz_i) \rho(\oz_j) \over \partial \oz_a}   {\partial  C(\oz,d) \over \partial \oz_b}  H^{-1}_{ab} +O(T)\ .
\end{eqnarray}
By using the definition of $C$ (Eq.~\ref{Eq:Cost_op}), we readily obtain Eq.~\ref{Eq:dFdWdbe}.

To calculate the $O(T)$ correction in Eq.~\ref{Eq:dFdWdbe} we either calculate the $O(T^2)$ term of Eq.~\ref{Eq:DevSP} or we derive Eq.~\ref{Eq:dFdW-2} with respect to $\beta$. Below we provide such a derivative for the four terms appearing in the $O(T)$ correction of Eq.~\ref{Eq:dFdW-2}: 
\begin{eqnarray}
{d H^{-1}_{ab} \over d \beta} &=& -\sum_{cd} H^{-1}_{ac} {d H_{cd} \over d \beta} H^{-1}_{db}
\nonumber \\
&=& -\sum_{cd} H^{-1}_{ac} \left[ {\partial H_{cd} \over \partial \beta} + \sum_e V^3_{cde} {\partial \oz_e \over \partial \beta} 
\right] H^{-1}_{db}\ ; \\
   {d \over d \beta} {\partial H_{ab} \over \partial W_{ij}} &=& \sum_c {\partial V^3_{abc} \over \partial W_{ij}} {\partial \oz_c \over \partial \beta}
\end{eqnarray}
where we have used that $f_{\theta,\alpha,\beta}$ is linear in $\theta$;
\begin{eqnarray}
    {d V^3_{abc} \over d \beta} = {\partial V^3_{abc} \over \partial \beta} + \sum_d  V^4_{abc}  {\partial \oz_d \over \partial \beta}\ ;
\end{eqnarray}
and
\begin{eqnarray}
    \sum_a H_{ka} {\partial^2 \oz_a \over \partial W_{ij} \partial \beta} = - \sum_a\left[ {\partial H_{ka} \over \partial W_{ij}} {\partial \oz_a \over \partial \beta} +{\partial H_{ka} \over \partial \beta} {\partial \oz_a \over \partial W_{ij}} \right] - \sum_{a,b} V^3_{kab} {\partial \oz_a \over \partial \beta} {\partial \oz_a \over \partial W_{ij}}
\end{eqnarray}
where we have taken the derivative of Eq.~\ref{Eq:doz_app} with respect to $\beta$, and
 $\partial \oz_a / \partial \theta $ is given by Eq.~\ref{Eq:dozdtheta}.

\bibliography{QTEquilProp}

\end{document}